\documentclass[conference,10pt]{IEEEtran}
\usepackage{epsfig,rotating,setspace,latexsym,amsmath,epsf,amssymb,amsfonts,bm,theorem,subfigure,epstopdf,cite,authblk, bbm}
\usepackage[ruled,vlined]{algorithm2e}

\usepackage{color}

\IEEEoverridecommandlockouts
\allowdisplaybreaks

\begin{document}

\title{Adversarial Machine Learning and Defense Game for NextG Signal Classification with Deep Learning}	
	
\author[]{Yalin E. Sagduyu}

\affil[]{\normalsize  National Security Institute, Virginia Tech, Arlington, VA, USA \\ Email: ysagduyu@vt.edu}
	
\maketitle
\begin{abstract}
This paper presents a game-theoretic framework to study the interactions of attack and defense for deep learning-based NextG signal classification. NextG systems such as the one envisioned for a massive number of IoT devices can employ deep neural networks (DNNs) for various tasks such as user equipment identification, physical layer authentication, and detection of incumbent users (such as in the Citizens Broadband Radio Service (CBRS) band). By training another DNN as the surrogate model, an adversary can launch an inference (exploratory) attack to learn the behavior of the victim model, predict successful operation modes (e.g., channel access), and jam them. A defense mechanism can increase the adversary's uncertainty by introducing controlled errors in the victim model's decisions (i.e., poisoning the adversary's training data). This defense is effective against an attack but reduces the performance when there is no attack. The interactions between the defender and the adversary are formulated as a non-cooperative game, where the defender selects the probability of defending or the defense level itself (i.e., the ratio of falsified decisions) and the adversary selects the probability of attacking. The defender's objective is to maximize its reward (e.g., throughput or transmission success ratio), whereas the adversary's objective is to minimize this reward and its attack cost. The Nash equilibrium strategies are determined as operation modes such that no player can unilaterally improve its utility given the other's strategy is fixed. A fictitious play is formulated for each player to play the game repeatedly in response to the empirical frequency of the opponent's actions. The performance in Nash equilibrium is compared to the fixed attack and defense cases, and the resilience of NextG signal classification against attacks is quantified.
\end{abstract}

\section{Introduction}
The \emph{next-generation (NextG) communication} systems are envisioned to rely on \emph{machine learning} to perform various complex tasks by learning from rich representations of spectrum data and adapting to spectrum dynamics. Backed up by recent computational and algorithmic advances, \emph{deep learning} has emerged as viable means to solve complex problems in NextG communications when conventional machine learning with hand-crafted features and analytical methods based on domain knowledge fall short from capturing the intrinsic characteristics of complex spectrum dynamics. 

\emph{Deep neural networks} (DNNs) can be effectively used to detect and classify wireless signals of interest \cite{erpek2020deep, dyspan} and have been successfully applied to various tasks such as user equipment identification, physical-layer user authentication, spectrum sensing and detection of incumbent users (e.g.,  radar signals in the CBRS band \cite{lees2019deep}).  As the spectrum resources are limited, they need to be effectively used by NextG communication systems. Although some frequency bands are originally dedicated to the use of incumbent users, the ever-growing demand for bandwidth raises the need to open these bands to NextG communications. A real-world example is the 3.5GHz Citizens Broadband Radio Service (CBRS) \cite{FCC}. The Federal Communications Commission (FCC) has opened the CBRS band (originally dedicated to incumbent users such as radar) to 5G and beyond communications. To enable the spectrum co-existence, spectrum sensors are deployed as IoT devices to form the Environment Sensing Capability (ESC) that collects RF signals to detect the incumbent users. The Spectrum Access System (SAS) configures 5G communications to prevent interference with the incumbent user. 

As the use of \emph{deep learning} in NextG communication system grows, a new security concern has emerged in the sense that the DNNs are susceptible  to the \emph{adversarial machine learning} attacks \cite{Sagduyu2020, Adesina2020, pajola}. Different wireless attacks built upon adversarial machine learning are possible including inference (exploratory) attacks \cite{Shi2018, Terpek, hou2019smart}, adversarial (evasion) attacks \cite{sadeghi2018adversarial, Flowers2020evaluating, kokalj2019targeted, Lin2020, Kim1, Kim2, 5GAML}, poisoning (causative) attacks \cite{YiMilcom2018, IoT2019, Sagduyu1, Luo3}, membership inference attacks \cite{MIA1}, spoofing attacks \cite{Shi2019generative}, Trojan attacks \cite{Davaslioglu1}, and covert communications \cite{Hameed2021the}.

In these attacks, the typical first step for the adversary is to build a \emph{surrogate model} of the victim system (in form of an \emph{inference attack}) and launch subsequent attacks using this trained model. Suppose that a NextG user (such as an IoT device deployed through the mMTC (massive Machine Type Communications) application of 5G and beyond network slices)  applies a DNN for signal classification. While the setting applies to other signal classification tasks such as signal authentication, spectrum sensing is used as the motivation example, where the goal of the DNN is to identify the spectrum opportunities and access the channel when no incumbent (background) transmission is detected. This way, high throughput and transmission (packet) success ratio can be achieved. In the meantime, an adversary can launch an inference attack, i.e, train a surrogate model, to predict when the channel is accessed. Then, the adversary can reliably predict when successful transmissions may occur and then jam these transmissions to reduce the spectrum access performance to a significant extent \cite{Terpek}. As a proactive \emph{defense} mechanism, it is possible to introduce controlled errors in spectrum access decisions (in form of a \emph{poisoning attack}) to poison the training data of the adversary, increase its uncertainty, and fool it into making wrong jamming decisions \cite{Terpek}. 

While this defense is effective against an adversary and helps the defender sustain its performance, it backfires and reduces the performance when there is no attack launched. The adversary may not be always active either for stealthiness or as a strategic decision to randomize its actions. The actions of the defender and the adversary are coupled together. The attack performance depends on whether the adversary launches an attack, or not. The impact of the attack depends on whether a defense is applied, or not, and furthermore depends on the level of this defense (namely, the ratio of controlled errors introduced in spectrum access decisions and consequently the ratio of poisoned training samples for the adversary). 

In this paper, these trade-offs are formulated as a \emph{non-cooperative game}. The effects of uncertainty have been considered for brute-force jamming games \cite{sagduyu2009MAC, Jammergame2, Garnaev1}. In this paper, the jamming attack is constructed to follow the inference attack  \cite{steal} (which is more effective than random or sensing-based jamming, as shown in \cite{Terpek}) and the interaction of the defense mechanism with the inference attack is studied as a game. First, the action of the defender is applying a fixed level of defense, or not. The action of the adversary is to launch an attack, or not. The strategies of the defender and the adversary are to select the probabilities of applying a defense or launching an attack, respectively. The defense aims to maximize its performance (throughput or transmission success ratio) as the reward. The loss of this performance is the reward for the adversary that also incurs the attack cost (for collecting spectrum sensing data, building surrogate model, and transmitting the jamming signal). Then, a \emph{fictitious play} is formulated, where each player plays the game repeatedly in response to the empirical frequency of the opponent's actions. Next, the game formulation is extended for the defender to select the defense level directly (instead of using a fixed level).

For all these game formulations, the \emph{Nash equilibrium} strategies are determined for the defender and the adversary such that no player can unilaterally improve its utility given the other's strategy is fixed. The gains and losses relative to the fixed cases of defense and attack are computed to quantify the effectiveness of the attack and defense mechanisms in the context of deep learning-based NextG spectrum. These results provide insights on the robustness of deep learning-based NextG signal classification against novel attacks. 

The rest of the paper is organized as follows. Section~\ref{sec:System} describes the system model. Section~\ref{sec:Game1} presents the game with probabilistic decisions of attack and defense (with fixed defense level). Section~\ref{sec:Game2} extends the game formulation with the selection of defense level. Section~\ref{sec:Conclusion} concludes the paper. 

\section{System Model for Attack and Defense} \label{sec:System}
\begin{figure}[t]
    \vspace{-0.5cm}
	\centering
	\includegraphics[width=\columnwidth]{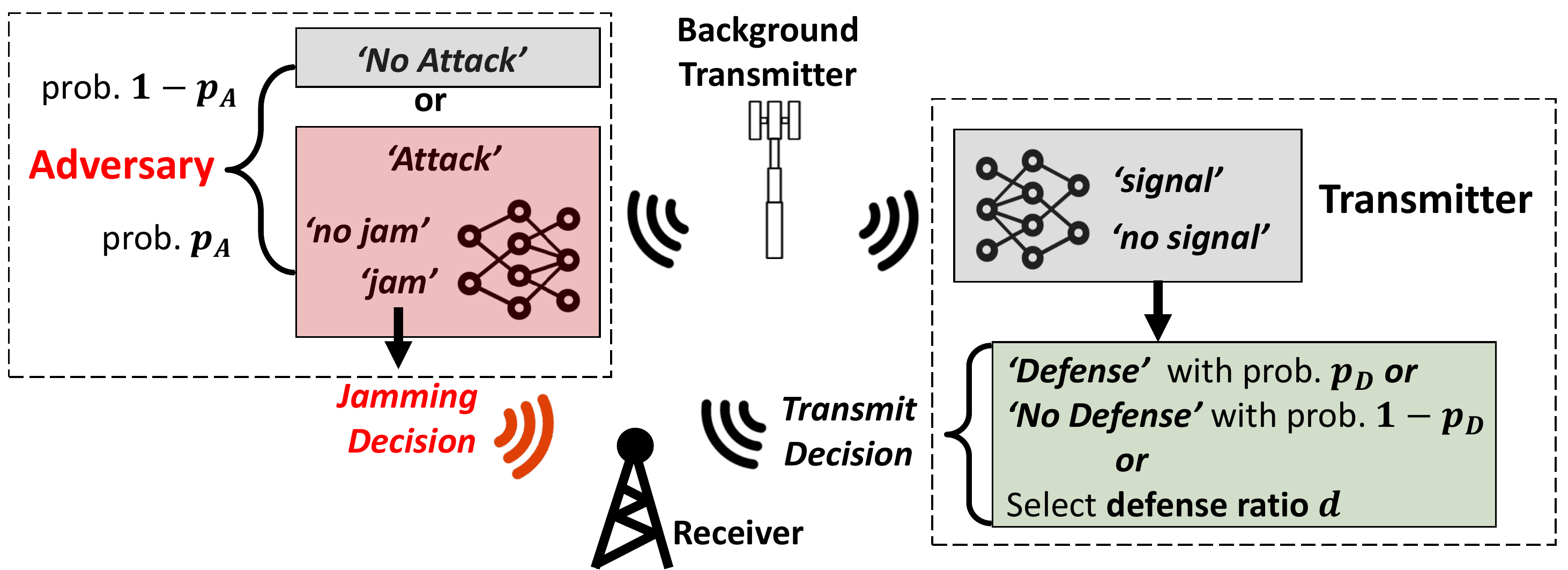}
	\caption{System model for attack and defense mechanisms.}
	\label{fig:system}
\end{figure} 
 The system model is illustrated in Fig.~\ref{fig:system}. There is one background transmitter (primary user) that chooses randomly at any given time to either transmit a signal with QPSK modulation, or remain idle. There is a NextG transmitter (secondary user) that senses the channel and transmits when there is no background transmission detected. For that purpose, a DNN, namely a convolutional neural network (CNN) is used to predict whether the channel is idle or occupied by the background transmission. 

The input to the CNN is of two dimensions (2,16) corresponding to 16 I/Q samples. The output labels are `signal' and `no signal'. Each client has 1000 data samples that are obtained from signals received over a Rayleigh fading channel, where Rayleigh distribution of parameter 1 is assumed and Additive White Gaussian Noise (AWGN) is added for 3dB signal-to-noise-ratio (SNR). The data samples are split into 80\% and 20\% portions to construct the training and test datasets, respectively. After varying hyperparameters, the architecture of the trained CNN model that yields the best accuracy of signal classification is shown in Table~\ref{table:DNN}). Categorical cross-entropy is used as the loss function and Adam is used as the optimizer. The numerical results are obtained in Python and the CNN model is trained in Keras with the TensorFlow backend. 

\begin{table}[h!]
	\caption{CNN architecture used for wireless signal classification.}
	\centering
	{\small
		\begin{tabular}{l|l}
			Layer & Properties \\ \hline \hline
			 Conv2D & filter size = 32, kernel size = (1,3),  \\ &  activation = ReLU \\ \hline
			 Flatten & -- 	 \\ \hline
			 Dense & size = 32, activation = ReLU	 \\ \hline
			 Dropout & dropout ratio =  0.1	 \\ \hline
			 Dense & size = 2, activation = Softmax	 \\ \hline
		\end{tabular}
	}
	\label{table:DNN}
\end{table}

There is one \emph{adversary} (i.e., a \emph{jammer}) that aims to predict when there would be a successful transmission of the NextG user (if there was no jamming) and jam only at those time instances. For that purpose, the adversary trains another CNN with the same architecture (but different weights) shown in Table~\ref{table:DNN}) as a \emph{surrogate model}. Another Rayleigh fading channel with AWGN at 3dB SNR is considered from the background transmitter to the adversary. The output labels are `successful transmission' (or `jam') and `no successful transmission' (or `no jam'). The NextG transmitter transmits if it predicts `no signal'. The classification accuracy is $0.9755$. If the background transmitter transmits or the adversary predicts `successful transmission' and jams, this transmission fails. Otherwise, it succeeds with some probability (assumed to be $0.95$). This jamming attack reduces the throughput (the ratio of successful transmissions by the NetxG transmitter over all time instances) from 0.4594 to 0.0532 (when the primary user is active half of the time) and the transmission success ratio (ratio of successful transmissions by the NextG transmitter over its all transmissions) from 0.8996 to 0.1091.

\begin{figure}[t]
    \vspace{-0.5cm}
	\centering
	\includegraphics[width=0.95\columnwidth]{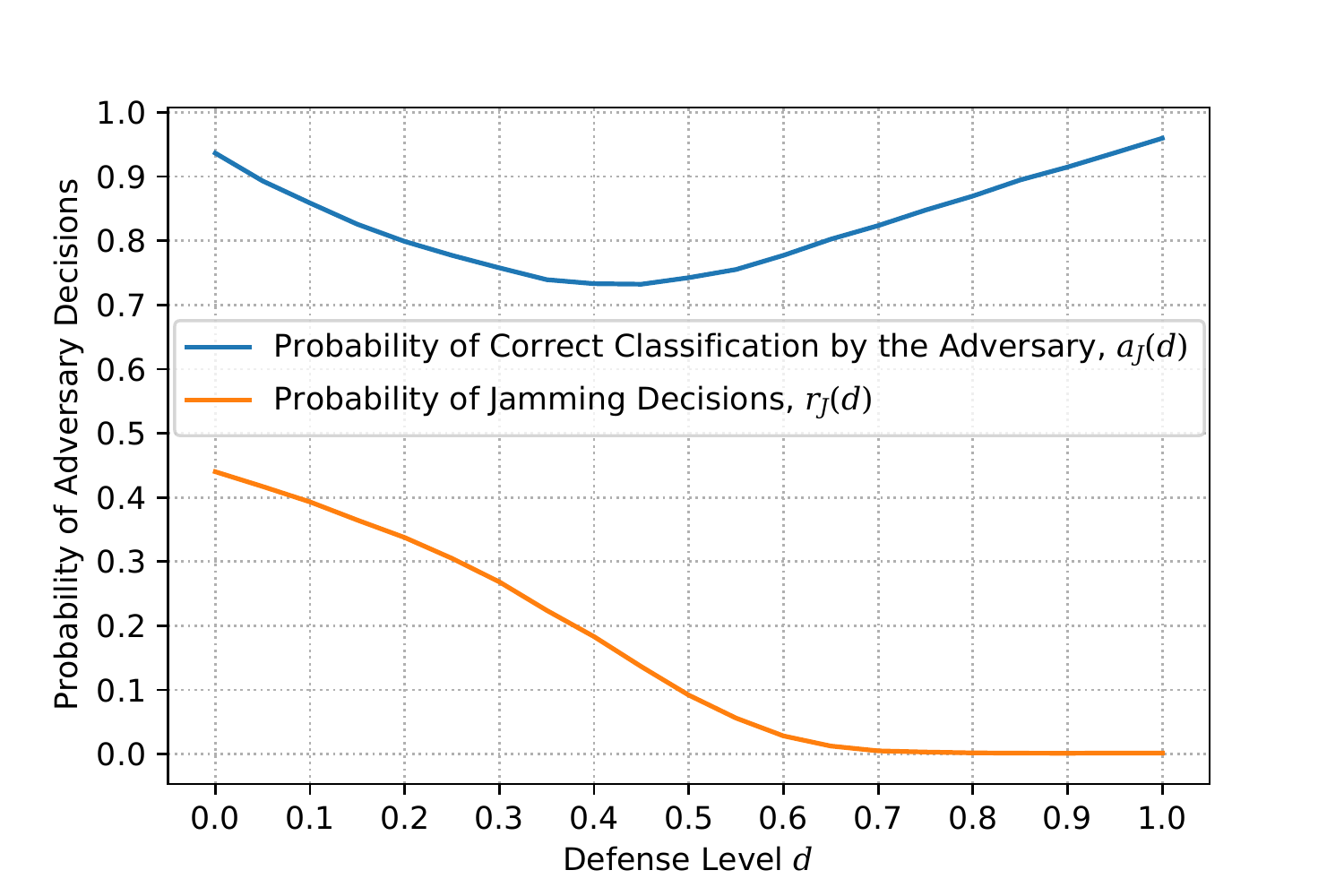}
	\caption{Probability of correct classification by the adversary, $a_J(d)$, and the probability of jamming decisions, $r_J(d)$.}
	\label{fig:jamratio}
\end{figure}

As a \emph{defense} mechanism, the NextG transmitter makes controlled errors in its transmission decisions such that the adversary cannot build a reliable surrogate model \cite{Terpek}. Defense level $d$ is defined as the ratio of spectrum access decisions flipped from the predicted label to the other label ($0 \leq d \leq 1$ and $d=0$ means no defense). For a given $d$, the decisions flipped are chosen at the time instances when the NextG transmitter makes its decisions with the highest confidence levels that are quantified by the Softmax output of the DNN  for each test sample. Fig.~\ref{fig:jamratio} shows the probability of correct classification by the adversary, $a_J(d)$, and the probability of jamming decisions, $r_J(d)$, as a function of $d$. Note that $a_J(d)$ drops first with $d$ (the adversary is fooled) but then increases with $d$ (large $d$ leads to few successful transmissions to predict), whereas $r_J(d)$ decreases with $d$ (the adversary has less tendency to launch a jamming attack as $d$ increases). 

\begin{figure}[t]
    \vspace{-0.5cm}
	\centering
	\includegraphics[width=0.95\columnwidth]{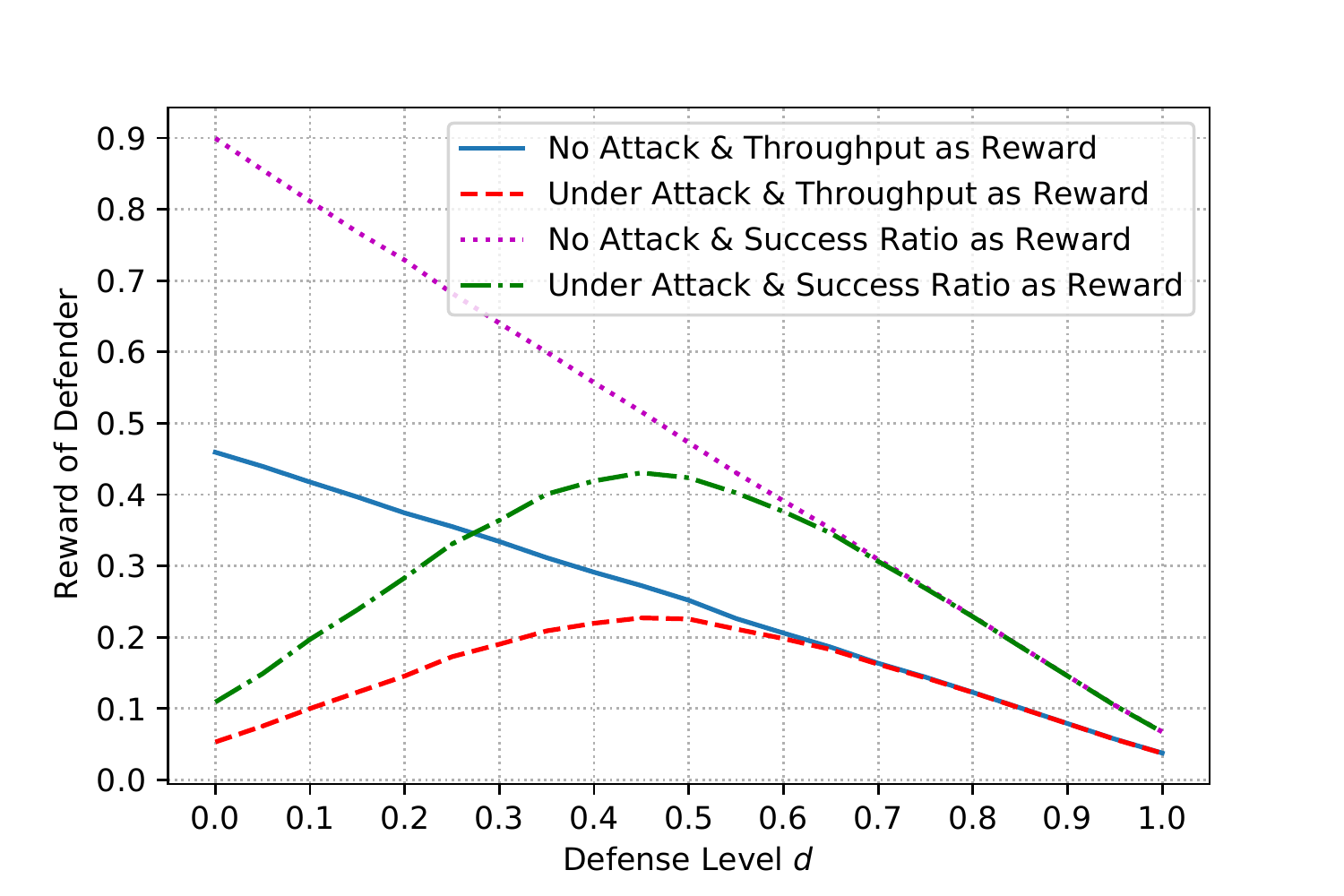}
	\caption{The reward of the defender as a function of the defense level $d$.}
	\label{fig:reward}
\end{figure}

Selecting a small $d$ may not be effective in terms of fooling the adversary, whereas a high $d$ hurts the transmitter's performance if there is no attack. 
Thus, there is an incentive for the defender not to apply a defense all the time. In the meantime, the adversary is more effective when it launches its attack when there is no defense applied. Therefore, there is an incentive for the adversary not to launch an attack all the time as it incurs a cost for attack (such as the cost for collecting spectrum sensing data, building the surrogate model, and jamming) and also needs to remain stealth. Thus, there are trade-offs for the decisions of the defender and the adversary such that the defender does not need to apply a defense all the time and the adversary does not need to launch an attack all the time. The performance of the defender is measured as its reward depending on throughput or success ratio. For defense level $d$, the reward is denoted by $u^{(D)}_{S_A}(d)$ when the adversary takes action $S_A \in \{A, \bar{A}\}$, namely $A$ for `attack' or $\bar{A}$ for `no attack'. The reward $u^{(D)}_{S_A}(d)$ as a function of $d$ is shown in Fig.~\ref{fig:reward}. $u^{(D)}_{\bar{A}}(d)$ decreases monotonically with $d$ (i.e., adding more defense always hurts the performance when there is no attack) and $u^{(D)}_{A}(d)$ first increases with $d$ (i.e., adding defense helps with the performance at the beginning by fooling the adversary) and then decreases with $d$ (i.e., adding more defense makes the spectrum access decisions unreliable).

\section{Game with Probabilistic Decisions of Attack and Defense} \label{sec:Game1}
The defender takes action $S_D \in \{D, \bar{D}\}$, namely $D$ for `defense' or $\bar{D}$ for  `no defense', and the adversary takes action $S_A \in \{A, \bar{A}\}$, namely $A$ for `attack' or $\bar{A}$ for `no attack'. To reflect randomized decisions, the strategies of the defender and the adversary are to select $p_D$ and $p_A$ as the probability of defender playing `defense' and the adversary playing `attack', respectively.  Given the opponent's action, let $u^{(D)}_{S_D|S_A}$ and $u^{(A)}_{S_A|S_D}$ denote the reward of the defender and the adversary, respectively, when the defender takes action $S_D \in \{D, \bar{D}\}$ and the adversary takes action $S_A \in \{A, \bar{A}\}$. When the defender plays `defense', it applies a fixed level of defense $d$, i.e., the rewards can be expressed as $u^{(D)}_{D|S_A} = u^{(D)}_{S_A}(d)$ and $u^{(A)}_{S_A|D} = u^{(A)}_{S_A}(d)$. The reward of the defender is either the throughput or the transmission success ratio.

The loss of the defender is the gain of the adversary, i.e., $u^{(A)}_{S_A|S_D} = - u^{(D)}_{S_D|S_A}$. The adversary can observe the defender's utility by observing the acknowledgments sent for successful transmissions. In general, $u^{(D)}_{D|A} > u^{(D)}_{\bar{D}|A}$ (i.e., the defense helps the defender when the adversary attacks) and  $u^{(D)}_{\bar{D}|\bar{A}} > u^{(D)}_{D|\bar{A}}$ (i.e., the defense backfires when the defender defends, namely adds controlled errors to its decisions, although there is no attack). The average utility of the defender in response to the adversary's strategy $p_A$ is given by
\begin{eqnarray} \label{eq:def}
u^{(D)} (p_D,p_A) \hspace{-0.3cm} & = & \hspace{-0.3cm} p_D \left( p_A u^{(D)}_{D|A} + (1-p_A) u^{(D)}_{D|\bar{A}}   \right) \\ && \hspace{-0.4cm} + (1-p_D) \left(  p_A u^{(D)}_{\bar{D}|A} + (1-p_A) u^{(D)}_{\bar{D}|\bar{A}} \right) \nonumber
\end{eqnarray}
and the average utility of the adversary in response to the defender's strategy $p_D$ is given by 
\begin{eqnarray} \label{eq:adv}
u^{(A)} (p_A, p_D) \hspace{-0.3cm} & = & \hspace{-0.3cm}p_A \left( p_D u^{(A)}_{A|D} + (1-p_D) u^{(A)}_{A|\bar{D}} - c_A\right) \\ && \hspace{-0.4cm} + (1-p_A) \left(  p_D u^{(A)}_{\bar{A}|D} + (1-p_D) u^{(A)}_{\bar{A}|\bar{D}}  \right), \nonumber
\end{eqnarray}
where $c_A$ is the cost of the attack (for collecting spectrum sensing data, building surrogate model, and jamming). 

\noindent {\bf Nash Equilibrium Strategies.} 
To compute the Nash equilibrium strategies,  $B_D\left(p_A\right)$ and $B_A\left(p_D\right)$ are defined as the best responses of the defender and the adversary, respectively, to each other's strategy. Then, the Nash equilibrium strategies satisfy $ p_D^* \in  B_D\left(p_A^* \right)$ and  $p_A^* \in  B_A\left(p_D^* \right)$,
where $B_D\left(p_A\right)$ and $B_A\left(p_D\right)$ are obtained by solving 
\begin{eqnarray} \label{eq:indvopt1}
\max_{p_i} \: \: u^{(i)} \left(p_i, p_{-i} \right) \:\:\: \text{subject to } 0 \leq p_i \leq 1,\label{eq:indvopt}
\end{eqnarray}
where $i \in \{D,A\}$ and $-i$ denotes the player other than $i$.
In Nash equilibrium, no player can unilaterally deviate from its strategy to increase its individual utility. To identify the Nash equilibrium strategies, the optimization in (\ref{eq:indvopt1}) is converted first to a standard form:
\begin{eqnarray}
\min_{x} f(x) \:\:\: \text{subject to } g_j(x) \leq 0, \label{eq:std}
\end{eqnarray}
where $f$ is the objective function, $x$ is the optimization variable, and $g_j$ is the $j$th inequality constraint function. The optimization in (\ref{eq:indvopt}) is converted to standard form (\ref{eq:std}) by setting 
\begin{eqnarray}
&& x = p_i, \label{eq:condfirst}\\
&& f(x) = - u^{(i)} \left(x, p_{-i}\right), \\
&& g_1(x) = -x, \\
&& g_2(x) = x - 1, \label{eq:condlast}
\end{eqnarray}
for given $p_{-i}$, where $g_1$ corresponds to the constraint $p_i \geq 0$ and $g_2$ corresponds to the constraint $p_i \leq 1$. In general, the Karush–Kuhn–Tucker (KKT) conditions for (\ref{eq:std}) are given by 
\begin{eqnarray}
&& \nabla_x f(x^*) + \sum_{j} \mu_{j} \nabla_x g_j(x^*) = 0, \label{eq:KKT2}\\
&& g_j (x^*) \leq 0, \:\:\:  \mu_j \geq 0, \:\:\: \sum_j \mu_j g_j(x^*) = 0 \label{eq:KKT3}.
\end{eqnarray}

By applying (\ref{eq:condfirst})-(\ref{eq:condlast}), the KKT conditions for the utility optimization for player $i \in \{D,A\}$ are given by
\begin{eqnarray}
&&  \hspace{-1.25cm} - \nabla_{p_i} u^{(i)} \left(p_i^*, p_{-i}\right) - \mu_{i,1} + \mu_{i,2} = 0, \label{eq:KKT1} \\
&& \hspace{-1.25cm} p_i^* \geq 0, \: p_i^* \leq 1, \: \mu_{i,j} \geq 0, \: \mu_{i,1}  p_i^* = 0, \: \mu_{i,1}  (p_i^*-1) = 0,
\end{eqnarray}
where $\mu_{i,j}$ is the KKT multiplier for the $j$th constraint of player $i \in \{D,A\}$.  (\ref{eq:KKT1}) couples strategies of the defender and the adversary, and can be expressed from (\ref{eq:def}) and (\ref{eq:adv}) as 
\begin{eqnarray} \label{eq:KKTspec1} 
&& p_A^* \left(u^{(D)}_{D|A} - u^{(D)}_{\bar{D}|A} +   u^{(D)}_{\bar{D}|\bar{A}} - u^{(D)}_{D|\bar{A}} \right) \\ && \hspace{0.55cm} = u^{(D)}_{\bar{D}|\bar{A}} - u^{(D)}_{D|\bar{A}} - \mu_{D,1} + \mu_{D,2}, \nonumber \\ \label{eq:KKTspec2} 
&& p_D^* \left( u^{(A)}_{A|D} - u^{(A)}_{A|\bar{D}} +   u^{(A)}_{\bar{A}|\bar{D}} - u^{(A)}_{\bar{A} |D} \right) \\ && \hspace{0.55cm} = u^{(A)}_{\bar{A}|\bar{D}} - u^{(A)}_{\bar{A}|D} - c_A - \mu_{A,1} + \mu_{A,2}, \nonumber
\end{eqnarray}
when $p_i^* \in [0, 1]$ for $i \in \{D,A\}$). Pure strategies ($p_i^* \in \{0, 1\}$ for $i \in \{D,A\}$) and mixed strategies ($p_A^*$ and $p_D^*$ obtained from (\ref{eq:KKTspec1})-(\ref{eq:KKTspec2}) with $\mu_{i,j} = 0$) may exist in Nash equilibrium.

\begin{figure}[t]
    \vspace{-0.5cm}
	\centering
	\includegraphics[width=0.95\columnwidth]{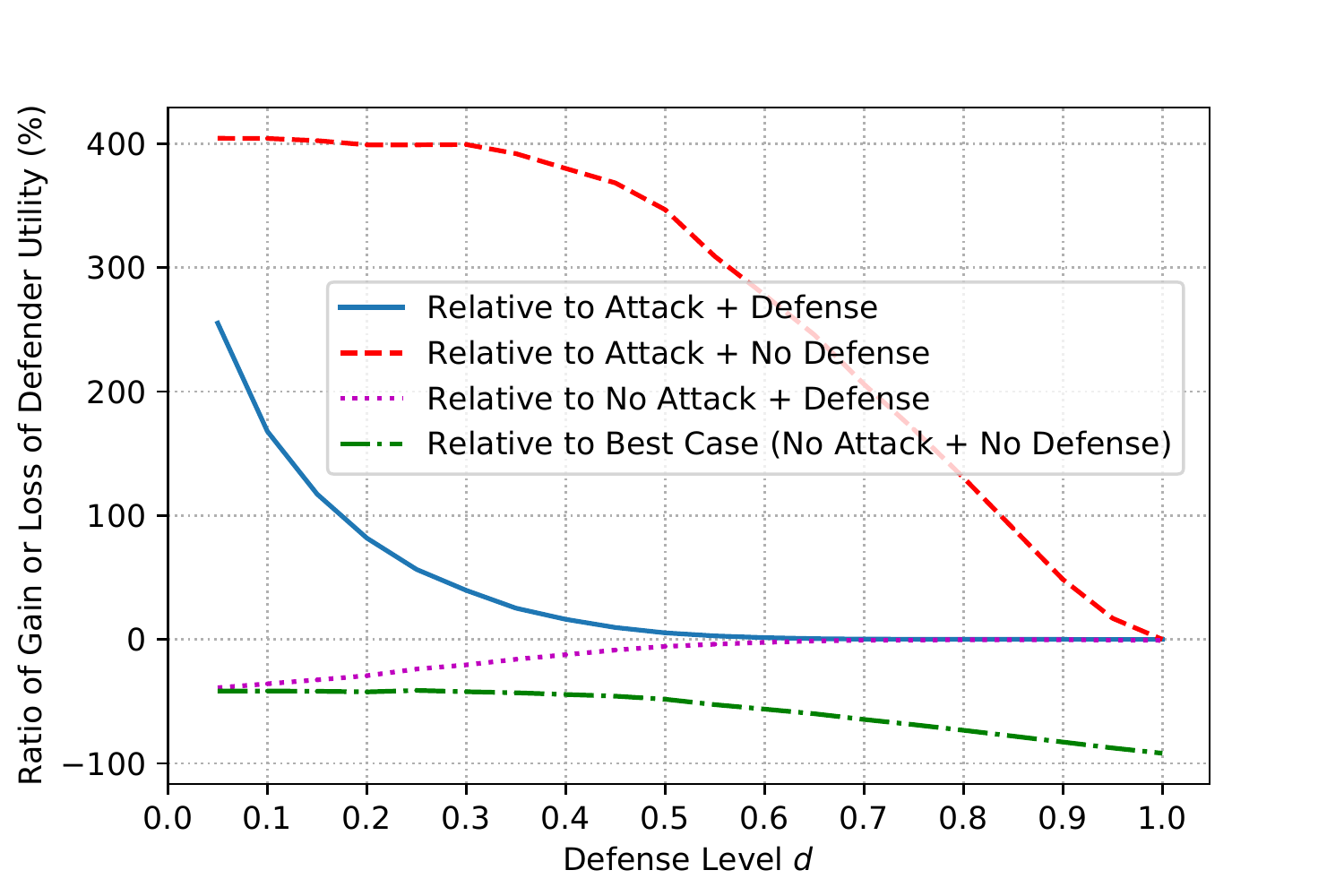}
	\caption{Ratio of gain or loss of the defender utility in Nash equilibrium with respect to different cases of attack and defense choices.} 
	\label{fig:ratios_univ}
\end{figure}

Next, the utility of the defender in Nash equilibrium is compared with different choices of fixed attack and defense strategies. Results that are averaged when the throughput and the success ratio are used for the reward are shown in Fig.~\ref{fig:ratios_univ}. The gains with respect to the case of `attack'+`defense' and the case of `attack'+`no defense' are $39\%$ and $269\%$, respectively, on average, and these gains decrease with the fixed level of defense $d$. The losses  with respect to the case of `no attack'+`defense' and the case of `no attack'+`no defense' are $12\%$ and $57\%$, respectively, on average, and decrease and increase, respectively, with $d$.

\begin{figure}[t]
    \vspace{-0.5cm}
	\centering
	\includegraphics[width=0.95\columnwidth]{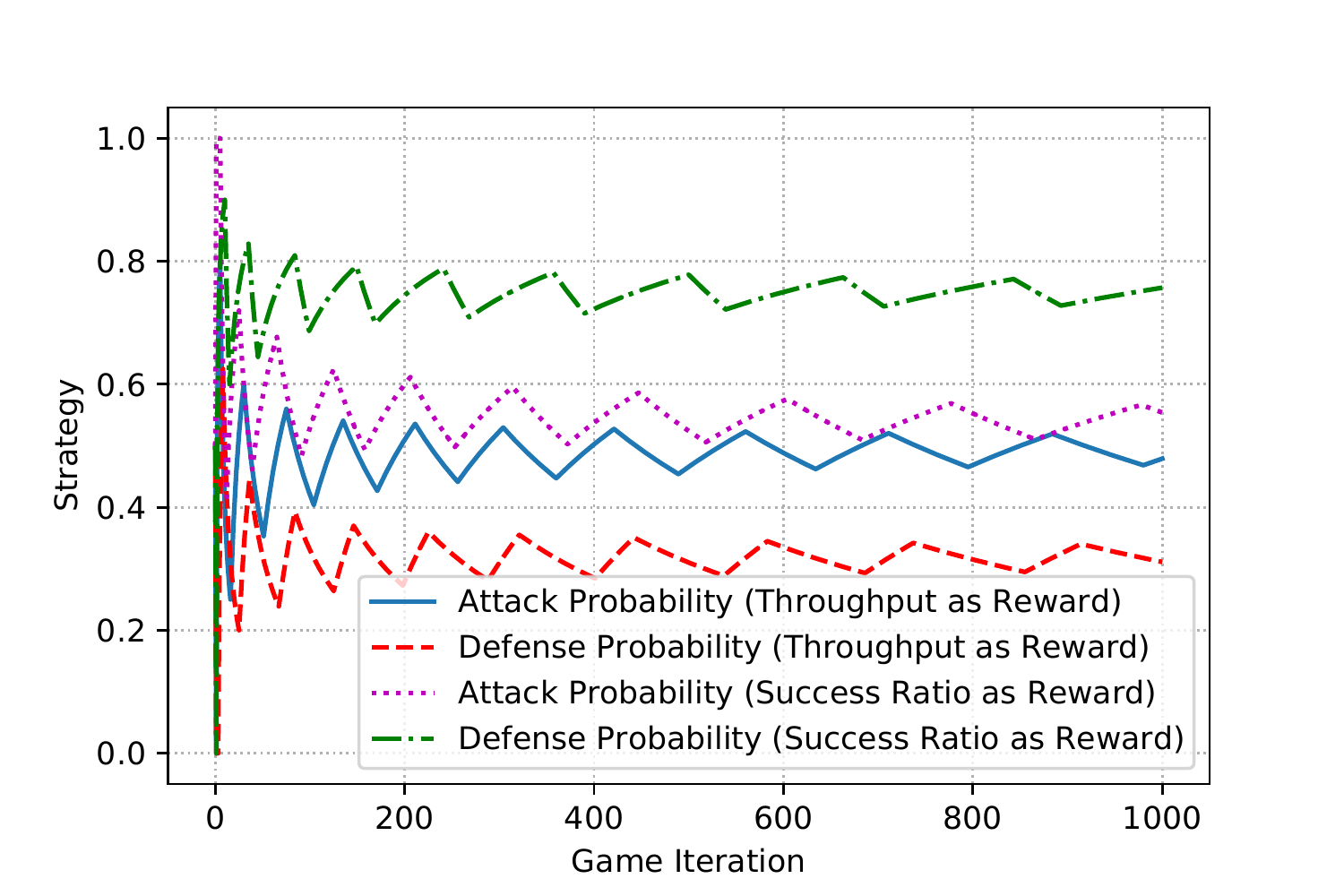}
	\caption{The empirical attack and defense probabilities learned over time.}
	\label{fig:fictplay1}
		    \vspace{-0.05cm}
\end{figure}

\begin{figure}[t]
    \vspace{-0.5cm}
	\centering
	\includegraphics[width=0.95\columnwidth]{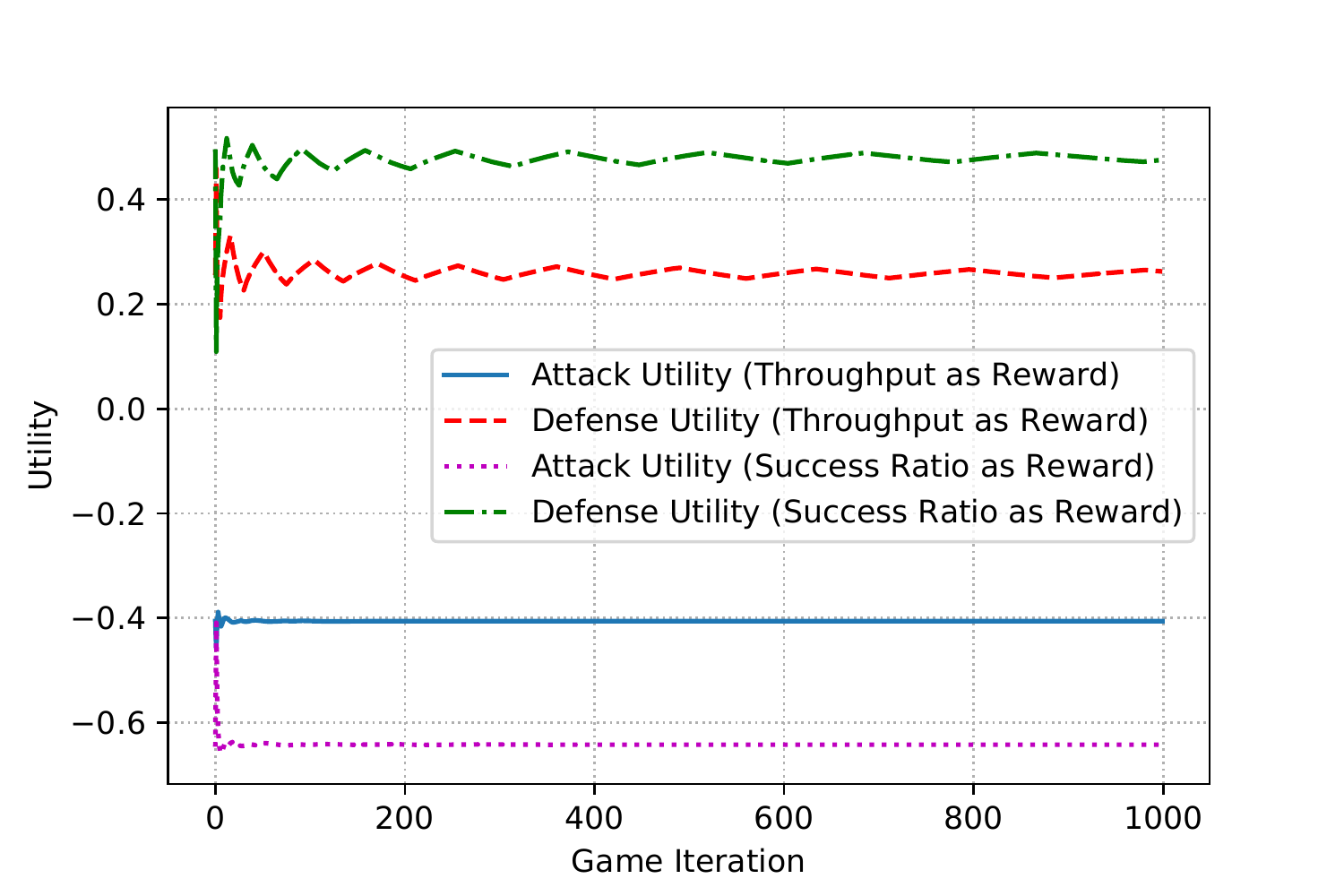}
	\caption{The empirical attack and defense utilities achieved over time.}
	\label{fig:fictplay2}
			   \vspace{-0.25cm}
\end{figure}

\noindent {\bf Fictitious Play.}  
Over multiple rounds, each player repeatedly plays the best response to the empirical frequency of the opponent's past actions without knowing each other's utility function. The defender observes the action of the adversary by computing its own reward, whereas the adversary observes the action of the defender by monitoring the acknowledgments sent for successful transmissions. Define $p_i(t)$ as the strategy assumed by player $i \in \{D,A\}$  at round $t$, and $\tilde{p}_{i}(t)$ as the moving average of the strategies played by player $i$ until round $t$. Then, player $i \in \{D,A\}$ plays strategy $p_{i}(t) = B_i(\tilde{p}_{-1}(t))$
at round $t$. Fig.~\ref{fig:fictplay1} and Fig.~\ref{fig:fictplay2} show how the strategies and the corresponding utilities of the defender and the adversary change over the game iterations of fictitious play (when $c_A = 0.3$ and $d = 0.4$). If both players select Nash equilibrium strategies at a given round, they continue with these strategies for subsequent rounds. If strategies of both players converge, these strategies constitute to a Nash equilibrium. 

\section{Game Extension with Defense Level Selection} \label{sec:Game2}
Instead of selecting between a fixed level defense or no defense, the defender can also select directly the level of defense $d$. The action of the adversary is `attack' or `no attack' such that $S_A \in \{A, \bar{A}\}$ and $p_A$ is the probability of launching an attack. On the other hand, the action of the defender is to select the defense level $d$ such that $S_D = d$. The reward of the defender is $u^{(D)}_{S_A}(d)$ and the loss of the defender is the gain for the adversary such that $u^{(A)}_{S_A}(d) = - u^{(D)}_{ S_A}(d)$. 

The average utility of the defender in response to the adversary's strategy $p_A$ is given by
\begin{eqnarray} \label{eq:def1}
u^{(D)} (d,p_A) =  p_A u^{(D)}_{A}(d) + (1-p_A) u^{(D)}_{\bar{A}}(d)   
\end{eqnarray}
and the average utility of the adversary in response to the defender's strategy $d$ is given by 
\begin{eqnarray} \label{eq:adv1}
u^{(A)} (p_A, d) \hspace{-0.2cm} & = & \hspace{-0.2cm} p_A \left( u^{(A)}_{A}(d) - c_{A,1} - c_{A,2} \: r_J(d) \right) \\ && \hspace{-0.2cm} + \left(1-p_A\right) u^{(A)}_{\bar{A}}(d) , \nonumber
\end{eqnarray}
where the cost of the adversary has two terms, $c_{A,1}$ is the fixed cost for collecting spectrum sensing data and building a surrogate model, $c_{A,2} r_J(d)$ is the cost transmitting the jamming signal, and $r_J(d)$ corresponds to the probability of jamming (as shown in Fig.~\ref{fig:jamratio}). The KKT conditions are 
\begin{eqnarray} \label{eq:spec3}
&& - \nabla_{d} \: u^{(D)} \left(d^*, p_A\right) - \mu_{D,1} + \mu_{D,2} = 0, \label{eq:KKT11} \\
\label{eq:spec4} && - \nabla_{p_A} u^{(A)} \left(d, p_A^*\right) - \mu_{A,1} + \mu_{A,2} = 0, \label{eq:KKT21} \\
&& d^* \geq 0, \: \: d^* \leq 1, \: \: p_A^* \geq 0, \:\: p_A^* \leq 1, \\
&& \mu_{D,1} \geq 0, \:\: \mu_{D,2} \geq 0, \:\: \mu_{A,1} \geq 0, \:\: \mu_{A,2} \geq 0, \\
&& \mu_{D,1} \: d^* = 0, \:\: \mu_{D,2} \: (d^*-1) = 0, \\ && \mu_{A,1} \: p_A^* = 0, \:\: \mu_{A,2} \: (p_A^*-1) = 0.
\end{eqnarray}

\noindent {\bf Nash Equilibrium Strategies.} From (\ref{eq:spec3}) and (\ref{eq:spec4}), the mixed strategies $d^* \in [0,1]$ and $p_A^* \in [0,1]$ in Nash equilibrium follow as solutions to
\begin{eqnarray}
&& p_A^* \left(-\nabla_d \: u_A^{(D)}(d^*) + \nabla_d \: u_{\bar{A}}^{(D)}(d^*)  \right)  \\ && \hspace{0.55cm}  = \nabla_d \: u_{\bar{A}}^{(D)}(d^*) + \mu_{D,1} - \mu_{D,2}, \nonumber
\end{eqnarray}
\begin{eqnarray}
&& - u^{(A)}_{A}(d^*) + c_{A,1} + c_{A,2} \: r_J(d^*) \\ && \nonumber  +  u^{(A)}_{\bar{A}}(d^*) -\mu_{A,1} + \mu_{A,2}=0.
\end{eqnarray}

\begin{figure}[t]
    \vspace{-0.5cm}
	\centering
	\includegraphics[width=0.95\columnwidth]{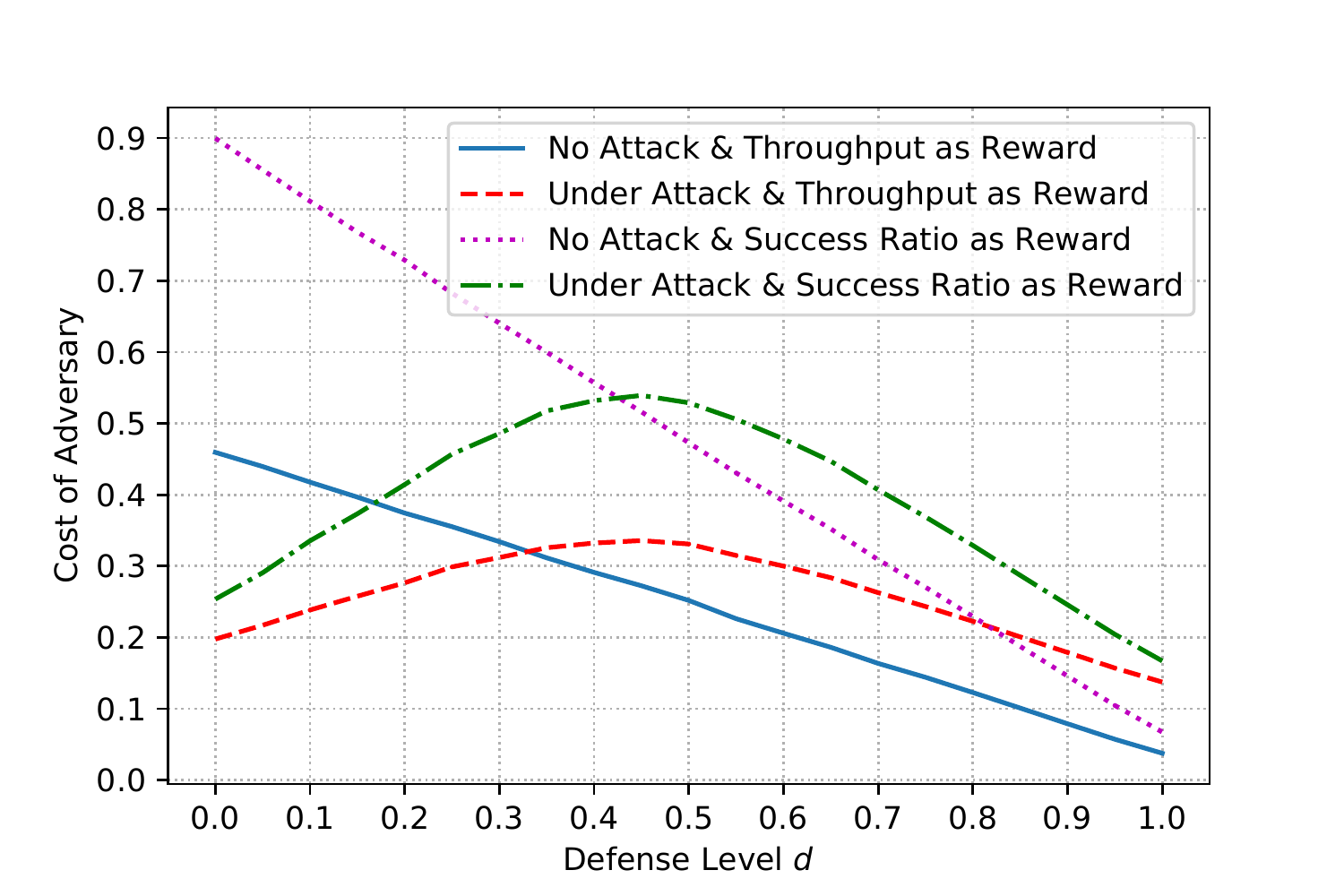}
	\caption{The cost of the adversary as a function of the defense level $d$.}
	\label{fig:crossing}
\end{figure}

Fig.~\ref{fig:crossing} shows the cost of the adversary as a function of the defense level $d$ for the two cases when the throughput or the success ratio is used as the reward (when $c_{A,1} = c_{A,2} = 0.1$). This cost is $u^{(D)}_{d|A} + c_{A,1} + c_{A,2} r_J(d)$, when the adversary decides to attack, and $u^{(D)}_{d|\bar{A}}$, when the adversary does not decide to attack. According to the KKT conditions, a mixed strategy of defense level $d^*$ in Nash equilibrium follows when the adversary selects $p_A^*$ such that the utility of the adversary is the same regardless of its action (namely, attack or not) and the adversary cannot unilaterally improve its performance. For example, from Fig.~\ref{fig:crossing}, $d^*$ can be derived when the costs of the adversary under the cases of attack and no attack intersect with each other. Then, $d^* = 0.33$ or $d^* = 0.42$ for $c_{A,1} = c_{A,2} = 0.1$, when the throughput or the success ratio is used, respectively, as the reward of the defender. 

\begin{figure}[t]
    \vspace{-0.5cm}
	\centering
	\includegraphics[width=0.95\columnwidth]{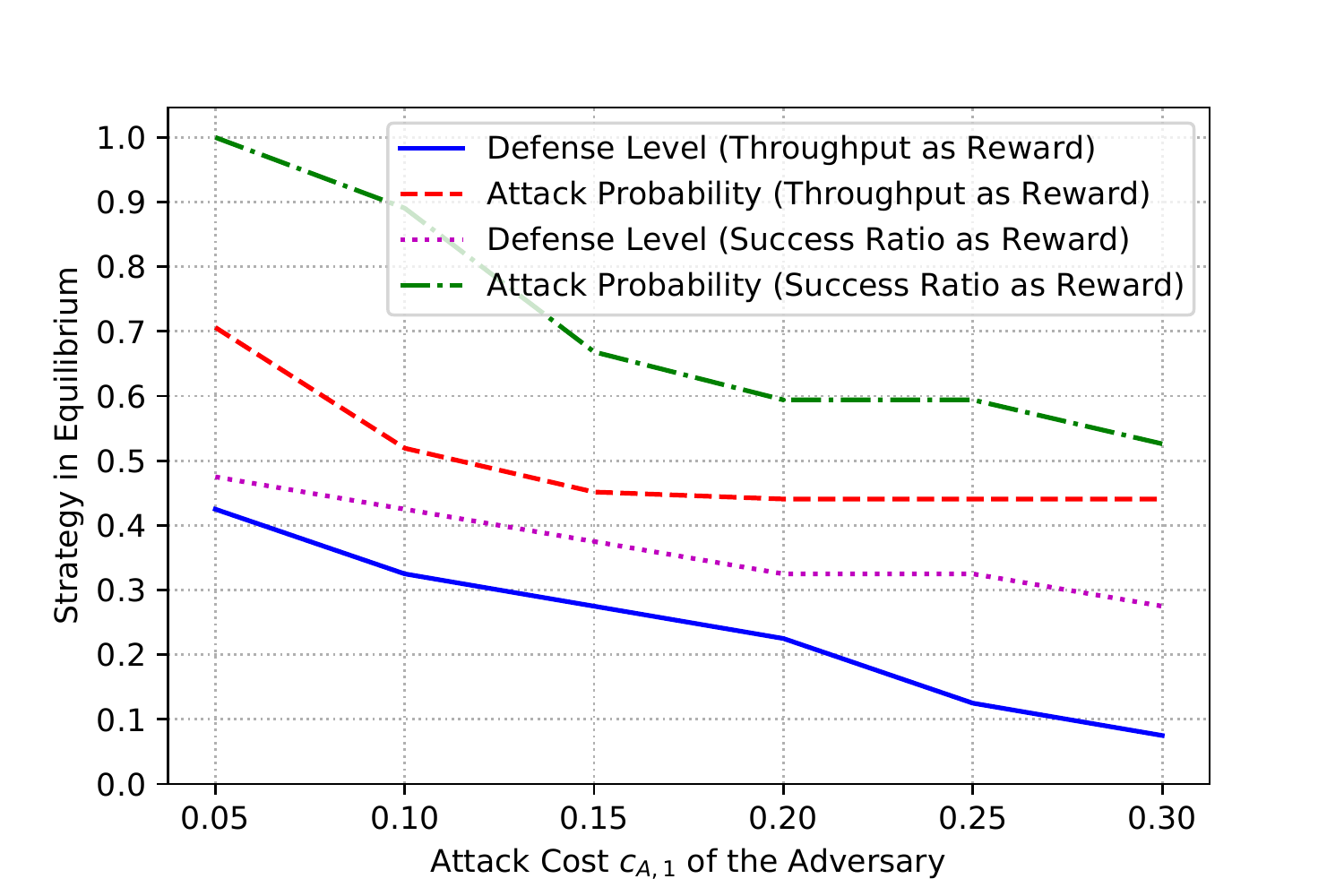}
	\caption{The equilibrium strategies as a function of the attack cost $c_{A,1}$.}
	\label{fig:peqdeq}
\end{figure}

\begin{figure}[t]
    \vspace{-0.5cm} 
	\centering
	\includegraphics[width=0.95\columnwidth]{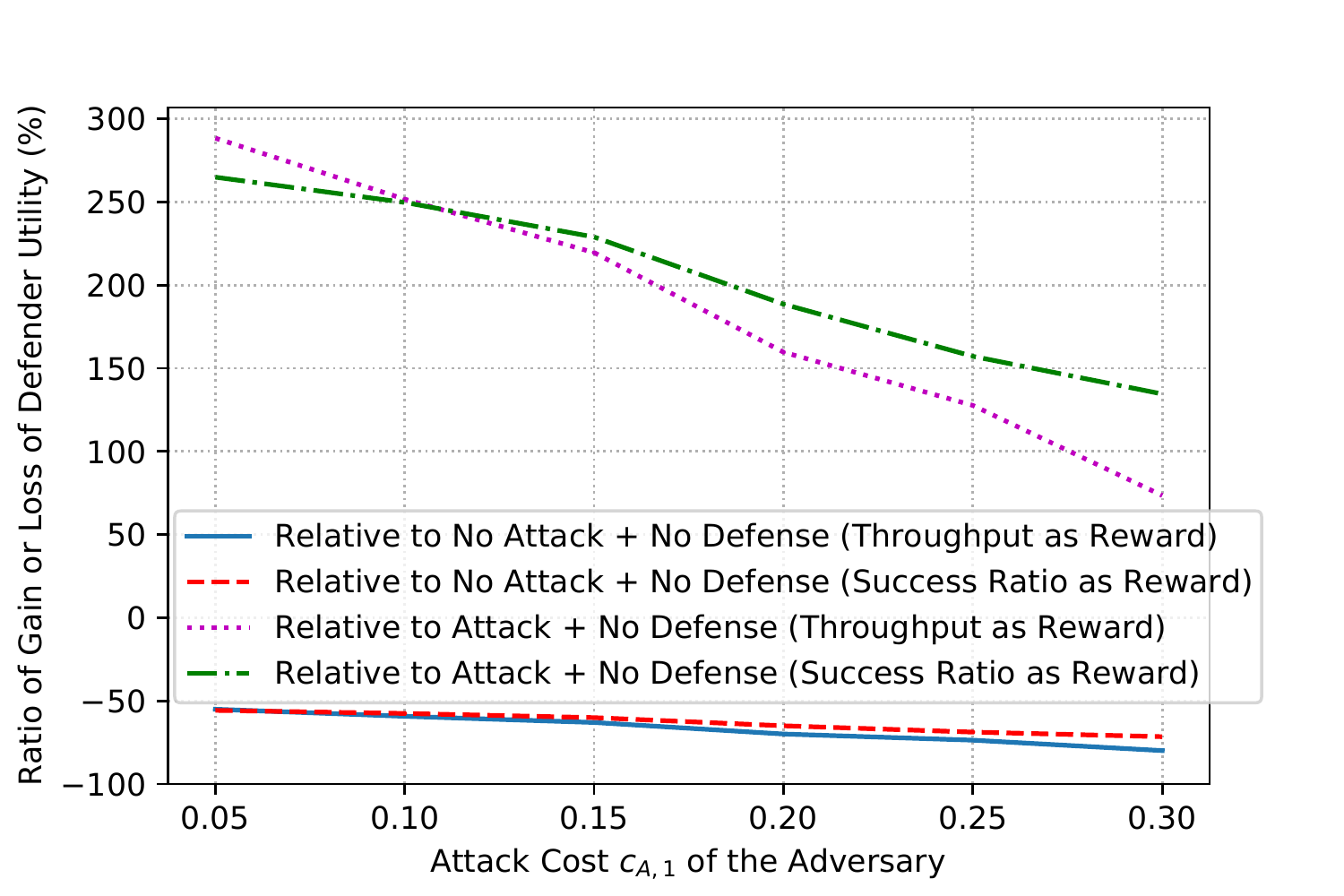}
	\caption{The average utility of the defender as a function of attack cost $c_{A,1}$.}
	\label{fig:defensereward}
\end{figure}

For $c_{A,2} = 0.1$, Fig.~\ref{fig:peqdeq} evaluates the Nash equilibrium strategies of the defender and the adversary ($d^*$ and $p_A^*$) as a function of $c_{A,1}$. Note that as $c_{A,1}$ increases, both $d^*$ and $p_A^*$ decrease, as the defender has less tendency to attack due to higher cost and therefore the defender can reduce its defense level. 
Fig.~\ref{fig:defensereward} shows the gains and losses of the Nash equilibrium strategies relative to the fixed cases of `attack' ($p_A = 1$) or `no attack' ($p_A = 0$) and `no defense' ($d = 0$). Results are obtained as a function $c_{A,1}$ for $c_{A,2} = 0.1$. In all cases, the relative performance in Nash equilibrium drops as the cost $c_{A,1}$ increases and the gains are much higher than the losses with respect to the fixed cases considered in Fig.~\ref{fig:defensereward}.  

\section{Conclusion} \label{sec:Conclusion}
A game-theoretic framework was presented to assess the attack and defense interactions for NextG signal classification, where spectrum sensors detect spectrum opportunities with deep learning. The first step of the attack is to build a surrogate model as an inference attack. The defense introduces errors into spectrum access decisions and poisons the adversary's training data to increase its uncertainty. The interactions of the defender and the adversary were formulated as a non-cooperative game, where the defender selects the probability of defending or the defense level, and the adversary selects the attack probability. The Nash equilibrium strafrformance relative to the fixed defense and attack cases was evaluated. Results quantified the attack impact and the defense resilience for deep learning-based NextG signal classification.           
\bibliographystyle{IEEEtran}
\bibliography{ref}

\begin{thebibliography}{10}
\providecommand{\url}[1]{#1}
\csname url@samestyle\endcsname
\providecommand{\newblock}{\relax}
\providecommand{\bibinfo}[2]{#2}
\providecommand{\BIBentrySTDinterwordspacing}{\spaceskip=0pt\relax}
\providecommand{\BIBentryALTinterwordstretchfactor}{4}
\providecommand{\BIBentryALTinterwordspacing}{\spaceskip=\fontdimen2\font plus
\BIBentryALTinterwordstretchfactor\fontdimen3\font minus
  \fontdimen4\font\relax}
\providecommand{\BIBforeignlanguage}[2]{{%
\expandafter\ifx\csname l@#1\endcsname\relax
\typeout{** WARNING: IEEEtran.bst: No hyphenation pattern has been}%
\typeout{** loaded for the language `#1'. Using the pattern for}%
\typeout{** the default language instead.}%
\else
\language=\csname l@#1\endcsname
\fi
#2}}
\providecommand{\BIBdecl}{\relax}
\BIBdecl

\bibitem{erpek2020deep}
T.~Erpek, T.~J. O’Shea, Y.~E. Sagduyu, Y.~Shi, and T.~C. Clancy, ``Deep
  learning for wireless communications,'' in \emph{Development and Analysis of
  Deep Learning Architectures}.\hskip 1em plus 0.5em minus 0.4em\relax
  Springer, 2020.

\bibitem{dyspan}
Y.~Shi, K.~Davaslioglu, Y.~E. Sagduyu, W.~C. Headley, M.~Fowler, and G.~Green,
  ``Deep learning for {RF} signal classification in unknown and dynamic
  spectrum environments,'' in \emph{IEEE International Symposium on Dynamic
  Spectrum Access Networks (DySPAN)}, 2019.

\bibitem{lees2019deep}
W.~M. Lees, A.~Wunderlich, P.~J. Jeavons, P.~D. Hale, and M.~R. Souryal, ``Deep
  learning classification of 3.5-{GHz} band spectrograms with applications to
  spectrum sensing,'' \emph{IEEE Transactions on Cognitive Communications and
  Networking}, 2019.

\bibitem{FCC}
``Citizens broadband radio service,'' \emph{Code of {F}ederal {R}egulations,
  Title 47, Part 96}, 2015.

\bibitem{Sagduyu2020}
Y.~E. Sagduyu, Y.~Shi, T.~Erpek, W.~Headley, B.~Flowers, G.~Stantchev, and
  Z.~Lu, ``When wireless security meets machine learning: Motivation,
  challenges, and research directions,'' \emph{arXiv:2001.08883}, 2020.

\bibitem{Adesina2020}
D.~Adesina, C.-C. Hsieh, Y.~E. Sagduyu, and L.~Qian, ``Adversarial machine
  learning in wireless communications using {RF} data: A review,'' \emph{IEEE
  Communications Surveys \& Tutorials}, 2022.

\bibitem{pajola}
L.~Pajola, L.~Pasa, and M.~Conti, ``Threat is in the air: Machine learning for
  wireless network applications,'' in \emph{ACM Workshop on Wireless Security
  and Machine Learning (WiseML)}, 2019.

\bibitem{Shi2018}
Y.~Shi, Y.~E. Sagduyu, T.~Erpek, K.~Davaslioglu, Z.~Lu, and J.~Li,
  ``Adversarial deep learning for cognitive radio security: Jamming attack and
  defense strategies,'' in \emph{IEEE International Communications Conference
  (ICC) Workshops}, 2018.

\bibitem{Terpek}
T.~Erpek, Y.~E. Sagduyu, and Y.~Shi, ``Deep learning for launching and
  mitigating wireless jamming attacks,'' \emph{IEEE Transactions on Cognitive
  Communications and Networking}, 2019.

\bibitem{hou2019smart}
T.~Hou, T.~Wang, Z.~Lu, and Y.~Liu, ``Smart spying via deep learning: inferring
  your activities from encrypted wireless traffic,'' in \emph{IEEE Global
  Conference on Signal and Information Processing (GlobalSIP)}, 2019.

\bibitem{sadeghi2018adversarial}
M.~Sadeghi and E.~G. Larsson, ``Adversarial attacks on deep-learning based
  radio signal classification,'' \emph{IEEE Wireless Communications Letters},
  2018.

\bibitem{Flowers2020evaluating}
B.~Flowers, R.~M. Buehrer, and W.~C. Headley, ``Evaluating adversarial evasion
  attacks in the context of wireless communications,'' \emph{IEEE Transactions
  on Information Forensics and Security}, 2020.

\bibitem{kokalj2019targeted}
S.~Kokalj-Filipovic, R.~Miller, and J.~Morman, ``Targeted adversarial examples
  against {RF} deep classifiers,'' in \emph{ACM Workshop on Wireless Security
  and Machine Learning (WiseML)}, 2019.

\bibitem{Lin2020}
Y.~Lin, H.~Zhao, Y.~Tu, S.~Mao, and Z.~Dou, ``Threats of adversarial attacks in
  {DNN}-based modulation recognition,'' in \emph{IEEE INFOCOM}, 2020.

\bibitem{Kim1}
B.~Kim, Y.~E. Sagduyu, K.~Davaslioglu, T.~Erpek, and S.~Ulukus, ``Over-the-air
  adversarial attacks on deep learning based modulation classifier over
  wireless channels,'' in \emph{Conference on Information Sciences and Systems
  (CISS)}, 2020.

\bibitem{Kim2}
------, ``Channel-aware adversarial attacks against deep learning-based
  wireless signal classifiers,'' \emph{IEEE Transactions on Wireless
  Communications}, 2022.

\bibitem{5GAML}
Y.~E. Sagduyu, T.~Erpek, and Y.~Shi, ``Adversarial machine learning for {5G}
  communications security,'' in \emph{Game Theory and Machine Learning for
  Cyber Security}, 2021.

\bibitem{YiMilcom2018}
Y.~Shi, T.~Erpek, Y.~E. Sagduyu, and J.~Li, ``Spectrum data poisoning with
  adversarial deep learning,'' in \emph{IEEE Military Communications Conference
  (MILCOM)}, 2018.

\bibitem{IoT2019}
Y.~E. Sagduyu, Y.~Shi, and T.~Erpek, ``{IoT} network security from the
  perspective of adversarial deep learning,'' in \emph{IEEE International
  Conference on Sensing, Communication and Networking (SECON) Workshops}, 2019.

\bibitem{Sagduyu1}
Y.~E. Sagduyu, T.~Erpek, and Y.~Shi, ``Adversarial deep learning for
  over-the-air spectrum poisoning attacks,'' \emph{IEEE Transactions on Mobile
  Computing}, 2021.

\bibitem{Luo3}
Z.~Luo, S.~Zhao, Z.~Lu, J.~Xu, and Y.~E. Sagduyu, ``When attackers meet {AI}:
  Learning-empowered attacks in cooperative spectrum sensing,'' \emph{IEEE
  Transactions on Mobile Computing}, 2022.

\bibitem{MIA1}
Y.~Shi, K.~Davaslioglu, and Y.~E. Sagduyu, ``Over-the-air membership inference
  attacks as privacy threats for deep learning-based wireless signal
  classifiers,'' in \emph{ACM Workshop on Wireless Security and Machine
  Learning (WiseML)}, 2020.

\bibitem{Shi2019generative}
------, ``Generative adversarial network for wireless signal spoofing,'' in
  \emph{ACM Workshop on Wireless Security and Machine Learning (WiseML)}, 2019.

\bibitem{Davaslioglu1}
K.~Davaslioglu and Y.~E. Sagduyu, ``Trojan attacks on wireless signal
  classification with adversarial machine learning,'' in \emph{IEEE
  International Symposium on Dynamic Spectrum Access Networks (DySPAN)
  Workshops}, 2019.

\bibitem{Hameed2021the}
M.~Z. Hameed, A.~György, and D.~Gündüz, ``The best defense is a good
  offense: Adversarial attacks to avoid modulation detection,'' \emph{IEEE
  Transactions on Information Forensics and Security}, 2021.

\bibitem{sagduyu2009MAC}
Y.~E. Sagduyu, R.~Berry, and A.~Ephremides, ``{MAC} games for distributed
  wireless network security with incomplete information of selfish and
  malicious user types,'' in \emph{International Conference on Game Theory for
  Networks}, 2009.

\bibitem{Jammergame2}
Y.~E. Sagduyu, R.~A. Berry, and A.~Ephremides, ``Jamming games in wireless
  networks with incomplete information,'' \emph{IEEE Communications Magazine},
  2011.

\bibitem{Garnaev1}
A.~Garnaev, A.~P. Petropulu, W.~Trappe, and H.~V. Poor, ``A jamming game with
  rival-type uncertainty,'' \emph{IEEE Transactions on Wireless
  Communications}, 2020.

\bibitem{steal}
Y.~Shi, Y.~Sagduyu, and A.~Grushin, ``How to steal a machine learning
  classifier with deep learning,'' in \emph{IEEE International Symposium on
  Technologies for Homeland Security (HST)}, 2017.

\end{thebibliography}

\end{document}